# Stability of monodomain III-V crystals and antiphase boundaries over a Si monoatomic step.

**D. Gupta[1], S. Pallikkara Chandrasekharan[1], S. Thébaud[1], C. Cornet[1,*], L. Pedesseau[1,*]**

[1]Univ Rennes, INSA Rennes, CNRS, Institut FOTON – UMR 6082, F-35000 Rennes, France

KEYWORDS Density functional theory, surface energy, interface energy, anti-phase boundaries, III-V/Si epitaxy.

**ABSTRACT:** Here, we compare the stabilities of different III-V crystals configurations on stepped Si substrates, with or without antiphase boundaries, for abrupt and compensated interfaces, using density functional theory. Thermodynamic stability of the different heterostructures is analyzed with an atomic scale description of charge densities distribution and mechanical strain. We show that the configuration where a III-V crystal adapts to a Si monoatomic step through change of charge compensation at the hetero-interface is much more stable than the configuration in which an antiphase boundary is formed. This study thus demonstrates that antiphase boundaries commonly observed in III-V/Si samples are not originating from Si monoatomic step edges but from inevitable kinetically driven coalescence of monophase 3D III-V islands.

# Introduction

Intense research efforts focused in recent years on heteroepitaxy of III-V semiconductors on silicon, because of its potential for the development of high quality and low-cost devices in the field of photonics[1,2,3,4] or energy applications.[5,6,7] Efficient devices demonstrations were made possible thanks to the recent efforts dedicated to the fundamental understanding of the physical processes involved during III-V/Si heteroepitaxy.[8–17] These works highlighted that most of the crystal defects, detrimental for devices and regularly encountered in III-V epilayers grown on Si, are generated at the very early crystal growth steps. More specifically, the impact of antiphase boundaries (composed of wrong III-III or V-V bonds in the III-V matrix) on optoelectronic properties of III-V/Si devices and materials was widely discussed. While this crystalline defect was considered for years as a device killer for lasers or photovoltaic solar cells,[3,18] it was also shown in recent works that they have original optoelectronic properties that may be appealing for devices provided their generation can be controlled.[7,13,19,20]

In the pioneering work of Kroemer[21], a simplified picture of their generation was proposed and adopted afterwards by most of the publications in the field, although Kroemer highlighted in his seminal work the probable higher complexity of the situation in real samples. In this picture, antiphase boundaries (APBs) are formed either by the presence of a monoatomic step at the Si surface that shifts the III-V lattice by half a monolayer and produces an APB or by the change of the group-III or group-V atoms bonded to the Si surface with an abrupt interface, as illustrated in Fig. 1. This picture is however in direct contradiction with recent experimental and theoretical works that led to the following unambiguous conclusions: (i) abrupt V-Si or III-Si interface are not stable. Instead, compensated III-V/Si interfaces are much more stable, independently of the chemical potential because of the better charge management at the heterogeneous interface.[14–16,22,23] (ii) the III-V/Si growth starts with monodomain Volmer-Weber islands,[9] where the III-V phase is entirely governed by the Si terrace on which it nucleates [11–13], indicating that the atomic configuration of the interface does not form randomly (iii) The size of 3D monodomain islands can be much larger that the distance between Si steps, so that the steps by themselves are not generating antiphase boundaries in the grown III-V crystal.[9] Instead, the substrate miscut can be useful to burry afterwards antiphase domains[11–13]. In all these studies, the generation of antiphase boundaries was only ascribed to the heterophase coalescence of individual monophase 3D III-V islands, but still without a clear description of the process at the atomistic scale and especially without any explanation on how an individual island can grow over a Si monoatomic step while remaining monodomain.

In this work, we compare the stability of III-V/Si heterostructures including a monoatomic step on the Si surface, with various atomic configurations, considering abrupt and compensated interfaces, with or without antiphase boundaries (APBs). The remarkable stability of structures accommodating the silicon monoatomic step without forming any antiphase boundary is discussed through a detailed analysis of mechanical forces and charges at the interface's level.

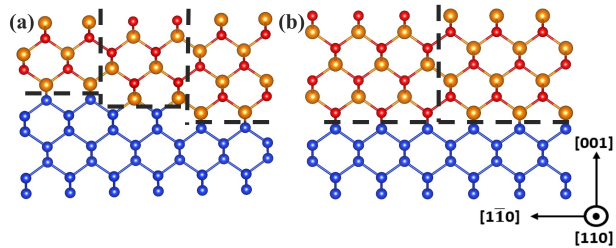

**Figure 1:** Historical view of APB formation during polar-on-nonpolar III-V/Si growth (a) with the presence of monoatomic steps on the substrate surface and (b) without the presence of monoatomic steps on the substrate surface. Recent experimental and theoretical findings contradict this view.[9,11–14]

# Computational details and simulated heterostructures

Here, the *ab initio* first principles calculations of different structures were performed by Spanish Initiative for Electronic Simulations with Thousands of Atoms (SIESTA[24,25]) within the density functional theory[26–28] (DFT) which is applied on bulk of Si, GaP with and without biaxial deformation (Table S1), and also heterostructures of GaP and Si materials. GaP is considered here, because of its low lattice mismatch with the Si. We calculated the energies of the bulk Si and bulk GaP in good accordance with the values obtained in our previous study[14]. However, during the epitaxial growth of GaP on silicon, the GaP undergoes a biaxial strain alongside [001] direction, as long as the layer thickness is below the critical thickness (*i.e.* <90 nm for GaP). Thus, the lattice parameter in the $[1\bar{1}0]$ and [110] direction remains the same as the one of silicon but varies in the [001] direction. The energy of the bulk GaP was then re-calculated using the modified lattice parameters respecting biaxial deformation as shown in Table S1. To maintain the homogeneity, the same computational parameters were used. Indeed, the biaxial deformation tends to destabilize the bulk GaP by 0.0308eV per Ga-P pair. The VESTA package[29] was used to build, and represent the heterostructures. To investigate the impact of APBs, 2 heterostructures with and without APBs and each with abrupt and compensated interfaces were constructed. According to ground state approximation of our DFT simulation, the entropic (T∆S) term is zero (T=0K), the p∆V term in the variation of enthalpy is also zero (volume is constant), and more the pressure does not vary. Thus, we can consider ∆U = ∆H = ∆F = ∆G = where U, H, F, and G are the internal energy, enthalpy, Helmholtz and Gibbs energies respectively. So, to be able to directly compare the energies obtained for each case, we imposed the same number of Si, Ga and P atoms with the same volume. The simulation of the 4 heterostructures was carried out using the generalized gradient approximation (GGA) functional in the Perdew-Burke-Ernzerhof (PBE)[30] form Troullier–Martins[31] pseudopotentials. The electronic wave functions are expanded onto a plane wave basis set with a cutoff energy of 150 Ry. A 1x1x1 Monkhorst-Pack grid[32] is used for the reciprocal space integration in the Brillouin zone. Heterostructures were minimized with the force criteria of 0.005 eV/Å by imposing the lattice parameter of the silicon on the GaP and releasing the GaP orthogonally to the silicon surface *i.e.* reaching a perfect biaxial deformation of the GaP.

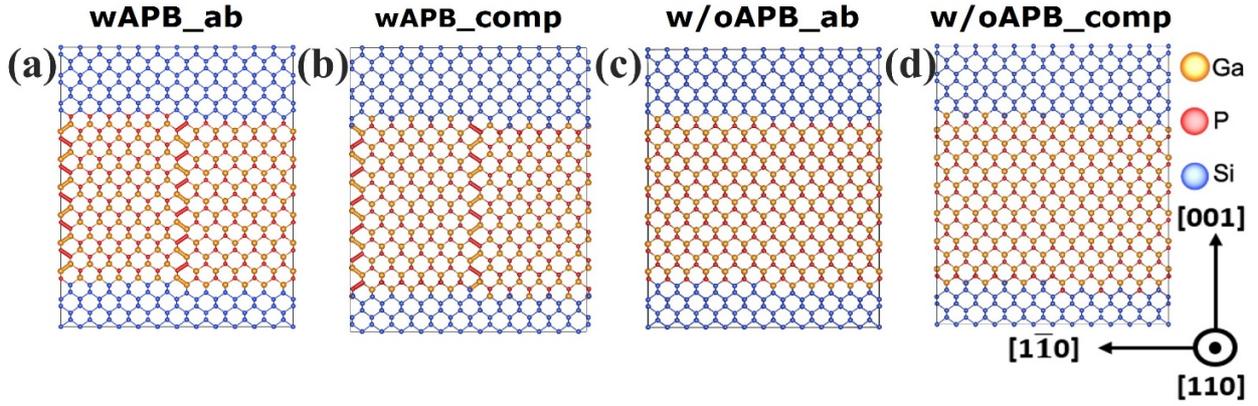

**Figure 2:** 2D representation of GaP/Si(001) heterostructures with a monatomic step at the III-V/Si interface. From left to right: (a) abrupt interface with APBs (b) compensated interface composed of mixed Si-Ga atomic rows with APBs (c) abrupt interface without APBs and (d) compensated interface with both mixed Si-Ga and Si-P interfacial layers without APB

The simulated heterostructures are shown in Figure 2. We model four different GaP/Si(001) heterostructures with abrupt or compensated interfaces, with or without APBs generated at the monoatomic Si step level. Fig. 2(a) represents the case where all the interfaces are perfectly abrupt, keeping always the same unique group-III or group-V atom bonded to the Si surface on the two sides of the monoatomic step. With this configuration, an APB is automatically formed at the step level. This configuration corresponds to the one presented in many papers in the literature and is represented in Fig. 1 (a) and will be referred as Abrupt interface with APBs (wAPBs_ab). Fig. 2(b) is similar to Fig. 2(a) except that all the interfaces are charge-compensated, with Ga and Si mixed atomic rows, corresponding to the most stable interfacial configuration for the GaP/Si interface, as already studied thoroughly. [14] As the compensation of the interface is the same on the two sides of the Si step, again, an APB naturally forms. This configuration is referred as Compensated interface with APBs (wAPBs_comp). Fig. 2(c) is similar to Fig. 2(a) except that abrupt interfaces are opposite on the two different sides of the Si monoatomic step, *i.e.* one is III-Si and the other one is V/Si. In this way, it becomes possible to adapt the silicon monoatomic step while preserving the III-V crystal ordering, and thus avoid the generation of APBs. This configuration is referred as Abrupt interface without APBs (w/oAPBs_ab). Fig. 2(d) is similar to Fig. 2(c) except that abrupt interfaces have been replaced by compensated ones. In this case, the compensation is different on the two different sides of the Si monoatomic step to preserve the III-V crystal ordering. On one side, the interface is composed of mixed Ga-Si atomic rows, while on the other side, the interface is composed of mixed P-Si atomic rows. With this strategy, APB is not needed to restore the crystal ordering. This configuration is referred as Compensated interface without APBs (w/oAPBs_comp).

On these four geometries, some atoms are frozen to mimic bulk materials and some are relaxed near the GaP/Si or APBs interfaces (Figure S1) to minimize the energy at the interface level. Each heterostructure is built with the same number of atoms for each element (384 Si, 288 Ga and 288 P) so that the energy computed can be compared directly and does not depend on the chemical potential. A smaller heterostructure (384 Si, 192 Ga and 192 P) is also built to mimic a smaller thickness of GaP, so that the results can be extrapolated and the energy evaluated for an infinitely small GaP thicknesses. In addition, the volume is also kept constant while respecting the response of a biaxial deformation of the GaP on the silicon substrate. To avoid Coulombic or dipolar interactions, the length between 2 steps is higher than 20Å[14] and the thicknesses of Si and GaP materials are larger than 25Å. In addition, when we build the heterostructures at the steps, the top and bottom steps are symmetric. Indeed, they are the same by applying two rotations by 180° each around [001] and around $[1\bar{1}0]$. The top and bottom interfaces are treated identically to decrease the error on the determination of the surface energies. The thicknesses of the four heterostructures are finally about 56Å ([001]) including 32Å of GaP.

## Results and Discussion

### (a) Contribution of excess energy to the total energy

We define the excess energy as a restatement of a previous study by Romanyuk *et al.*[15]

$$E_{excess} = E_{tot} - \Sigma_i \eta_i \mu_i \qquad (1)$$

where $E_{tot}$ is the total energy of the slab, $\eta_i$ is the number of the atoms of the species i, and $\mu_i$ is the chemical potential of the species i. $E_{excess}$ is thus mainly the excess energy corresponding to the contribution of the heterogeneous III-V/Si interfaces and/or APBs (if any) to the total energy (GaP/Si in our case). Although local strain may arise and contribute to this excess energy, originating from the non-infinite size of the slab, or from the non-perfect positioning of the frozen atoms, these effects remain neglectable in this study.

To determine the Excess energy, the 4 heterostructures (Fig. 2) were relaxed, as described in Fig. S1 at the same constant volume accounting for the small inherent biaxial deformation since the heterostructure is constrained to the Silicon lattice parameter, except along the [001] axis. Specifically, to accurately mimic the bulk, only two atomic layers plus two more atomic layers on each side of the heterogeneous III-V/Si interface and of the APBs were allowed to relax as shown in figure S1. The rest of the Si and GaP atoms are frozen on their positions, and are considered in this work as pseudo-bulk atoms. After the relaxation of the four heterostructures, the total energy was extracted, and, using equation (1), the excess energy was finally determined and reported in Table 1. The smaller the excess energy, the more stable the configuration is. Values are also given for the slabs with smaller numbers of atoms. From Table 1, the general trend can be easily captured. Configurations with APBs are less stable than configurations w/o APBs, and configurations with abrupt interfaces are less stable than configuration with compensated ones.

First, the better stability of the heterostructures with compensated interfaces is consistent with previous studies. Indeed, compensated interfaces unlike abrupt ones consist of an intermixing of Si-Ga or Si-P atoms at the interfacial layer. As shown in previous works, the most energetically stable interfaces are the ones composed of Si/Ga or Si/P,[14,22] with a ratio 0.5/0.5.[33] These interfaces unlike abrupt ones fulfill the electron counting model (ECM) which results in a lower interface energy due to their high stability. Here, we note that, although both Si-Ga and Si-P compensated interfaces are stable, the 0.5:0.5-Ga:Si is the most stable one. [14,22] For the slabs discussed here, the total energy is reduced by around 5 eV when the abrupt interface is replaced by an interface compensating the charges for a similar structure. Second, the fact that the presence of an APB in a slab destabilizes the system is also not surprising, as APBs are made of wrong bonds inside the perfect Zinc-Blende crystal. For the slabs discussed here, the total energy is increased by around 10 eV when an APB is inserted in the heterostructure. The atomic configuration (abrupt interfaces and APBs formation induced by the steps) chosen in fig. 2a corresponding to the historical picture depicted in Fig. 1a thus appears as a very unstable configuration and is thus unlikely to be promoted during the epitaxial III-V/Si growth.

As the total number of wrongly bonded atoms in APBs depend on the III-V thickness deposited, one can wonder is this conclusion remains true at the very beginning of the growth (for a very small thickness of III-V semiconductor deposited).

To confirm the consistency of our results with 960 atoms, the same calculations were done but with fewer (768) atoms (384 Si, 192 Ga and 192 P) and the results are given in parentheses in Table 1. The conclusion remains unchanged.

**Table 1 Excess energy of the biaxial strained heterostructures are considered here, with or without APBs, with compensated or abrupt interface.**

|  | wAPB_ab | wAPB_comp | w/oAPB_ab | w/oAPB_comp |
|---|---|---|---|---|
| Number of atoms | 960 | 960 (768) | 960 (768) | 960 (768) |
| Excess energy, $E_{excess}$ (eV) | 191.67 | 186.93 (148.69) | 181.14 (147.14) | 175.65 (140.93) |

Figure 3 represents the excess energy as a function of the GaP thickness expressed in monolayers (ML) for the w/oAPB_comp and wAPB_comp atomic configurations with respectively 8 ML and 12 ML. Solid and dashed lines represent the extrapolation of the Excess energy to the very beginning of the growth (0 ML of GaP). The inset gives a zoomed version of the graph for small GaP thicknesses. In this zoomed figure, it can be seen that when the GaP thickness tends to 0, the w/oAPB_comp configuration remains 0.6 eV more stable that the wAPB_comp one. Although this quantitative estimation may be subjected to uncertainties because of residual local stress, it confirms again the very low stability of APBs at the early stage of growth. This result not only confirms that APBs are very likely formed only during the out-of-equilibrium heterophase coalescence of monophase islands, but also explains the impressive faceting observed at the surface at the direct vicinity of emerging APBs, at the beginning of the growth[10] or for thicker III-V layers deposited on Si[11]. In other words, during III-V/Si heteroepitaxy, APBs are so unstable that the formation of energetically costly non (001) facets are preferred, rather than incorporating atoms with wrong bond configurations. The surface energy of non (001) facets being different for the various III-V semiconductors, the kind of faceting developed at the surface near the emerging APBs also depends on the material's system considered, as suggested by Gilbert *et al*.[13] [13]

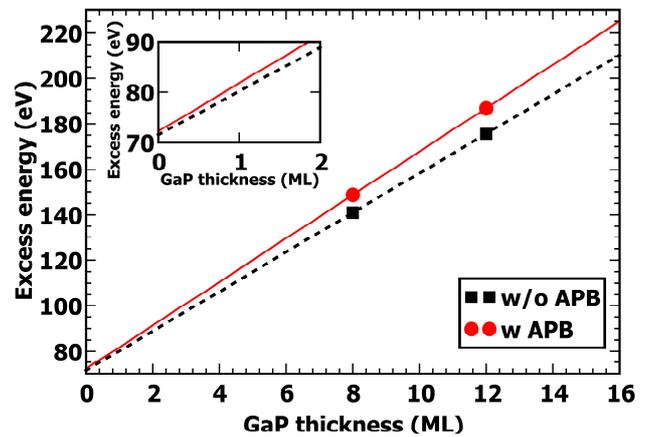

**Figure 3 Excess energy as a function of GaP thickness in monolayers (ML) with and without APBs**

In the following section, in order to clarify the main origin of the energetic instabilities induced by abrupt interfaces and APBs, electronic and mechanical properties of the studied heterostructures are analyzed thoroughly.

### (b) Electronic Properties

Here, we discuss the profile of the internal electric field in the heterostructures. The Si material is intrinsically nonpolar, while GaP is a polar material. In the heterostructures described previously, the local deviation to the natural stoichiometry induced by the chemical mismatch (heterogeneous interfaces, or APBs), may lead to an inherent electric field. Figure 4 represents the profiles of the total charge density, in electron/Å$^3$, that are averaged in the 2D plane ([110], [001]) as a function of the distance along the $[1\bar{1}0]$ direction, in Å, for different heterostructures (w/oAPB_ab, wAPB_comp, and w/oAPB_comp). These three structures have been selected to analyze separately the influences of APBs and abrupt interfaces on the charge distribution. For all heterostruc-tures, the red (white) color is used for the Step (Terrace) area from 0 to 46Å along the $[1\bar{1}0]$ direction.

First, at the terrace level (far away from the Si step), charge profiles are somehow comparable for the three structures. Indeed, in this area, the lateral averaging only reveals the expected charges of bulk GaP and Si, *i.e.* a neutral material overall. The only small (but significant) difference is that the charges fluctuations are less pronounced for the w/oAPB_comp configuration, which means that the charge sharing seems more equilibrated when the interface is compensated, and without the presence of Antiphase boundaries. Secondly, the most important difference between the different heterostructures is observed at the step level (red area). Indeed, the w/oAPB_ab configuration exhibits distinctive charge density peaks and valleys coinciding exactly with the position of the monoatomic step. The shift from a positive charge density to a negative one is the clear signature of a dipole forming at this level. For this specific configuration, the presence of pure Si-Ga bonds on one side of the step, and pure Si-P bonds on the other side of the step are expected to play a crucial role in the charge density variations. This effect completely disappears with the w/oAPB_comp configuration, that allows to conclude that the charge compensation of the interface is very effective to homogenize the electric field in the structure. Finally, we note that only small charge density fluctuations originate from APBs even at the step level, as seen for the wAPB_comp. This is due to the presence of the same number of Ga-Ga and P-P bonds because {110}-APB are stoichiometric overall, considering that the lateral averaging performed does not allow to see some more local charge fluctuations. Thus, the profile of the wAPB_comp is in reality the result of a compensation of total charge densities between Ga-Ga and P-P bonds. Interestingly, the color gradient highlights the dissipation of the atomistic reorganization (section: Mechanical properties) around the step and the APB. Indeed, we observe that the localized states caused by atomic rearrangements of the APB induce a non-local distortion which is discussed in the next section.

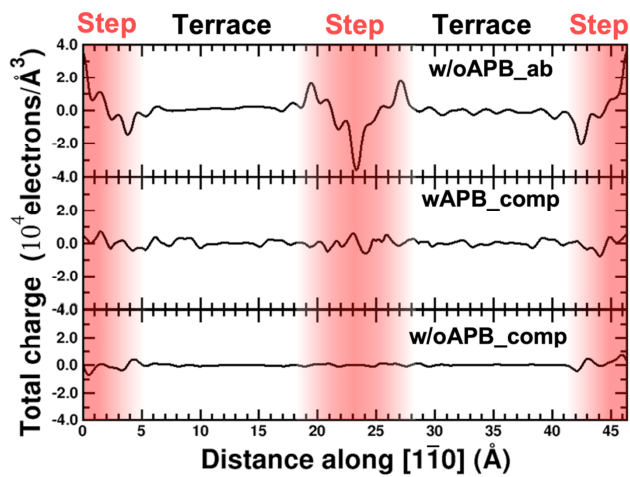

**Figure 4** Laterally and macroscopically averaged profiles of total charge density along $[1\bar{1}0]$

### (c) Mechanical properties

In general, for semiconductor materials, inter-mixing of two different atoms in a single atomic layer or the presence of struc-tural defect such as the formation of wrong bonds in a heterostructure can strongly impact mechanical properties, lattice ordering and strain in the heterostructures. This, in turns, can significantly change the optoelectronic properties and thermodynamic stability of the material. More specifically, when atoms of a semiconductor move as a response to an applied force, their local displacement may produce a local strain. In the present work, the applied force can be related to the atomic reorganization around the hetero-interface with III-V and silicon atoms but also around structural defect such as {110}-APBs. Subsequently, these atoms located immediately around the interface apply an internal force in the GaP semiconductor as well as an external force in the silicon lattice causing the local environment to become shifted from its original position. In this study, we do not consider the possible secondary phases SiP, SiP$_2$, GaSiP$_3$, GaSi$_3$, GaSi, and Ga$_3$Si that are in competition with both bulk Si and GaP, because the GaP-Si interface was shown to be stable (see ref. [14] Fig. S8 and Table SII).

In the following, we discuss and analyze the forces present at the GaP/Si interface and at the {110}-APBs for the different heterostructures. To do so, we compare the initial position of atoms before relaxation, corresponding to a perfect biaxially strained crystal, and the position of atoms after the whole relaxation process. The displacement along one direction is assumed to be proportional to the differential force field ΔF along that direction (before and after the relaxation). Figure 5 (a) shows the differential force field (on each atom) along the $[1\bar{1}0]$ direction for the wAPB_ab configuration. The atomic lattice is represented in the $[1\bar{1}0]$ and [001] coordinate system, similarly to Fig. 2. Positive (negative) differential forces correspond to green (red) dots on the image, implying that any atom represented in green (red) is shifted toward the left (right) boundary of the image during the relaxation process. Area of the circles used at the atomic positions are proportional to the intensity of the differential force applied on the corresponding atom. As shown in Figure 5 (a), the differential force field along the $[1\bar{1}0]$ direction is clearly stronger at the vicinity of the APB. It is interesting to observe that the Ga-Ga differential force intensity is stronger than the P-P one. These results can be explained by looking at bulk phases of Ga and P. Indeed, from ref. [14] Table II, α-Ga phase and Black Phosphorus have bond length of 2.83Å and 2.28Å respectively. Consequently, the bond length differences, as compared to the GaP bulk are 17% and 5% respectively. Thus, the bond length variation is clearly greater for the α-Ga phase which can explain such a stronger differential force for Ga-Ga wrong bonds, as compared to P-P ones. In addition, the direction of the forces associated with the wrong Ga-Ga or P-P bonds are at the opposite from each other along the $[1\bar{1}0]$ axis (except in the area near the top and bottom steps, but this is associated to side effects due to the complex atomic arrangement near the step and the interaction between APB atoms and steps atoms). Overall, a clear mechanical strain is evidenced in the bulk III-V region, near the APB. However, the deviation to the bulk bonds behavior is very localized. Indeed, atoms that are not involved in the wrong bonds have their positions shifted by 0.1% at the maximum, as compared to the reference bulk GaP.

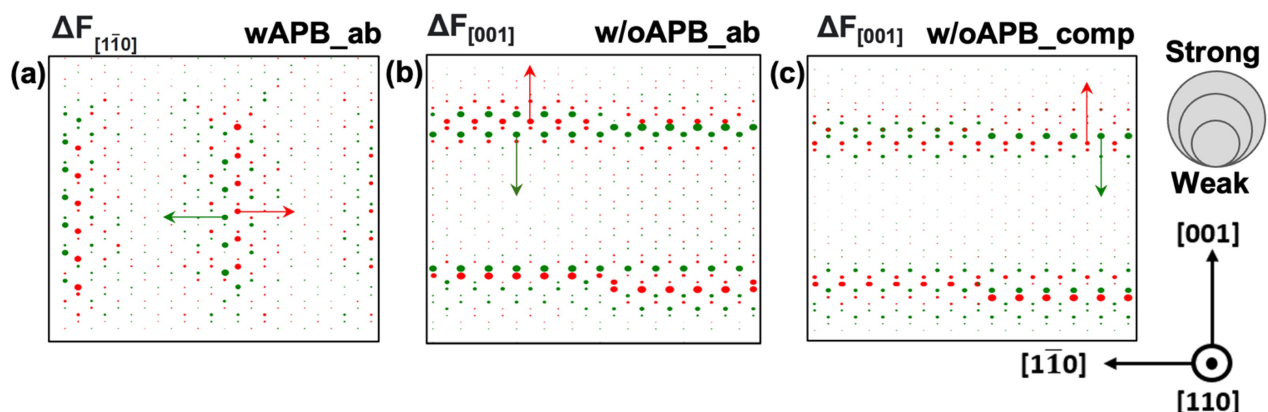

Figure 5 Differential Force Field (a) along $[1\bar{1}0]$ for the wAPB_ab configuration, (b) along [001] for the w/oAPB_ab configuration and (c) along [001] for the w/oAPB_comp configuration.

In Figure 5 (b) and (c), the differential force fields are represented for the w/oAPB_ab and w/oAPB-comp heterostructures, along the [001] direction, which is perpendicular to the III-V/Si heterointerfaces. Along that direction, the strongest differential forces are observed for the atoms positioned near the interface (terrace+step). Similarly to the previous case (Fig. 5 (a)), a mechanical strain is observed in the III-V region, induced by the III-V/Si interface. But again, this strain field impacts 4 to 5 rows of atoms and vanishes rapidly. However, the intensity of the differential forces is significantly stronger in the w/oAPB_ab (Fig. (b)) case than in the w/oAPB_comp (Fig. (c)) case. This is consistent with the fact that w/oAPB_comp (w/oAPB_ab) respects (does not respect) the ECM as shown in section (a), hence has lesser (higher) strain and higher (lesser) stability. Indeed, Figure 5(c) shows the differential forces of the most stable heterostructure, the forces are much smoother and decrease more rapidly than the other heterostructures. Overall, we show that atomic configurations at a III-III, V-V or III-V/Si heterointerface strongly impact the charge density sharing, which has drastic consequences on the mechanical strain and the stability of the system.

Conclusion

In conclusion, we have used DFT calculations to compare quantitatively the stabilities of GaP layers on stepped Si substrates, with or without antiphase boundaries, for abrupt or compensated interfaces. Especially, the configuration where a III-V APB is located at the vertical of a Si monoatomic step is found to be thermodynamically relatively unstable, while the configuration where a Si monoatomic step is adapted without APB, through a change of the interface compensation, appears much more stable. This conclusion is supported by a detailed analysis of charge densities and mechanical properties of the studied heterostructures. This study thus reveals that APBs are intrinsically highly unstable features in a III-V crystal, and can be formed only during the kinetically-driven inevitable coalescence of monophase III-V islands. It also shed some light on the appearance of high index facets observed at the surface, near emerging APBs, in III-V/Si bi-domain samples. The clarification of APBs generation processes demonstrated in the present work, is a real change of paradigm for the understanding of group-III-V/group-IV epitaxy, opening ways to better control defects generation in photonic or energy devices.

ASSOCIATED CONTENT

- Supporting Information

The Supporting Information is available free of charge at https://pubs.acs.org/doi/XXXX.

Table of stability of bulk Si and GaP with and without biaxial deformation; and 2D representation of GaP/Si(001) heterostructure with a monatomic step at III/V interface depicting relaxed and constrained atoms (PDF)

AUTHOR INFORMATION

Corresponding Author

*E-mail: charles.cornet@insa-rennes.fr; laurent.pedesseau@insa-rennes.fr

Author Contributions

C. C. and L. P. conceived the idea and designed the simulation, D. G. and L. P. did the DFT simulation, D. G., S. P. C., S. T., C. C. and L. P. analyzed the data, D. G., C. C. and L. P. wrote the paper. All authors reviewed and commented on the manuscript.

Funding Sources

This research was supported by the French National Research NUAGES (Grant no. ANR-21-CE24-0006) and PIANIST (Grant No. ANR-21-CE09-0020) projects.

ACKNOWLEDGMENT

This research was supported by the French National Research NUAGES (Grant no. ANR-21-CE24-0006) and PIANIST (Grant No. ANR-21-CE09-0020) projects. DFT calculations were performed at Institut FOTON, and the work was granted access to the HPC resources of TGCC/CINES/IDRIS under the allocation 2022-A0120911434 and 2023-A0140911434 made by GENCI.

ABBREVIATIONS

APB, antiphase boundary; DFT, density functional theory; ECM, electron counting model; ML, monolayer.